\documentstyle[12pt,aps,prd,epsfig,preprint]{revtex}

\begin{document}
\draft
\def\beq{\begin{equation}}
\def\eeq{\end{equation}}
\def\bea{\begin{eqnarray}}
\def\eea{\end{eqnarray}}
\def\ba{\begin{array}}
\def\ea{\end{array}}
\def\nnb{\nonumber}
\def\ve{\vert}
\def\vel{\left|}
\def\ver{\right|}
\def\ga{\left(}
\def\dr{\right)}
\def\aga{\left\{}
\def\adr{\right\}}
\def\rar{\rightarrow}
\def\la{\langle}
\def\ra{\rangle}
\def\lla{\left<}
\def\rra{\right>}
\def\ds{\displaystyle}
\def\0{\c{s}}   

\title{\bf The influence of the mutual drag of carrier-phonon system
on the thermopower and the transverse Nernst-Ettingshausen effect}
\author{M. M. Babaev$^{a}$, T. M. Gassym$^{a,b}$, M. Ta\0$^{b}$
\thanks{e-mail:~tasm@metu.edu.tr}~ and M. Tomak$^{b}$}
\address{$^a$Institute of Physics, Academy of Sciences of Azerbaijan
Baku 370143, Azerbaijan\\
$^b$Physics Department, Middle East Technical University 06531
Ankara, Turkey}
\maketitle
\begin{abstract}
The thermopower and Nernst-Ettingshausen (NE) effect in degenerate 
semiconductors and semimetals placed in high electric and magnetic 
fields are calculated by taking into account the heating of both 
electrons and phonons as well as their thermal and mutual drags.

The magnetic and electric field dependences of the thermoelectric
power and the transverse NE voltage are found in analytical forms.
It is shown that in weak and high transverse magnetic fields, the
electronic and phonon parts of NE coefficients change their sign
for some scattering mechanisms.  
\end{abstract}  
\thispagestyle{empty}
~~\\  
\narrowtext
\newpage
\setcounter{page}{1}

\section{Introduction}
The theoretical and experimental interest in thermoelectric power
in bulk and recent low dimensional systems has been
intensified\cite{1}-\cite{13}. A relatively long survey of literature
and some common misunderstandings in the field of thermoelectric
power ($\alpha$) and Nernst-Ettingshausen (NE) effect under different
transport conditions\cite{3,7,8,9},\cite{15}-\cite{23} are given in
our recent paper\cite{Kane}.

Lei\cite{11} showed in 1994 that the diffusion component of~ $\alpha$~
may be negative at a low lattice temperature range and high electric
field while the phonon drag component is still positive. Such a result
was also obtained by Babaev and Gassymov\cite{20} in 1977. They
theoretically investigated the NE effect and~ $\alpha$~ in semiconductors
at high electric and nonquantizing magnetic fields by solving the coupled
system of kinetic equations for electrons and phonons. The electron and
phonon heating, and the phonon drag were taken into account. It was shown
that when the temperature gradient of hot electrons is produced only by
the lattice temperature gradient, the electronic parts of the
thermoelectric and NE fields reverse their sign. In the case of phonon
heating and~ ${\ds T_p=T_e \gg T}$, both electronic and phonon parts of
the thermoelectric and thermomagnetic fields reverse their sign for all
cases considered. Here~ $T_e$, $T_p$ and $T$~ are the temperature of
electrons, phonons and lattice, respectively.

The NE effect and $\alpha$ in II-VI semiconductors have been investigated
with increasing interest\cite{31}-\cite{34}. The earlier investigations
of the magnetic field ($H$) dependence of the longitudinal NE effect in
HgSe\cite{35,36} and lead chalcogenides\cite{37,38} in the region of
higher temperatures ($T \geq 77 K$) demonstrated that the
thermoelectromotive force exhibits saturation in the region of strong
magnetic fields irrespective of the dominant scattering mechanism of 
charge carriers in the conduction band. However, the longitudinal NE 
effect in iron-doped HgSe samples at low temperatures ($20 \leq T \leq 
60~K$) has a maxima in the plot of $\Delta \alpha(H)=\mid \alpha(H)-
\alpha(0) \mid$. $\Delta \alpha(H)$ first increases quadratically with
increasing $H$ for $\Omega \tau <1$ then passes through a maximum at
$H=H_m$, and finally decreases as the field increases further (here 
$\Omega=eH/mc$ is the cyclotron frequency and $\tau$ is the electron
relaxation time). Another unusual fact is the sign reversal of the
transverse NE coefficient $Q_{\perp}(H)$ with increasing $H$ in the range
$\Omega\tau>1$\cite{33,34}. The experiments in gallium-doped HgSe
revealed that at low temperatures the NE coefficients change sign with
increasing gallium concentration or the applied magnetic field strength. 
These unusual features of the NE effect may be attributed to the effect 
of mutual drag which can be observed in semiconductors with high
concentration of conduction electrons\cite{39}. 

In the absence of external magnetic field, the $\alpha$ of hot electrons, 
taking into account the heating of phonons and the thermal drag, is
considered in Ref. 18. In that paper, the deformation potential of
interaction between electrons and phonons is considered. The transverse
NE effect and $\alpha$ of hot electrons in nondegenerate semiconductors
are studied in Ref. 40 without taking into account the effect of phonon
drag and their heating; and in Ref. 19 by taking into account the thermal
drag only in transverse magnetic field. However, these studies did not
consider the mutual drag of charge carriers and phonons.

There are some investigations considering the electron-phonon drag
and transport phenomena in semiconductors\cite{report}-\cite{Mash3}.
In Ref. 41, the electron-optical phonon drag and the size effect are
mainly considered. The Refs. 43 and 44 also considered the size
effect in finite semiconductors under the conditions of mutual drag.
The electric current and electron and phonon parts of the thermal
fluxes are obtained in general forms for the degenerate statistics
of electrons in Ref. 42. Gurevich and Mashkevich list the procedure
for determining the distribution function of electrons. However,
the list is not complete. Because, they obtain only the general
expressions for electric current and electron and phonon parts of
thermal fluxes, but they did not find the external electric field
dependence of the effective electron and phonon temperatures.
Therefore, they did not find the thermoelectric coefficients and
their external electric field dependence.

In the present paper, the NE effect and $\alpha$ in degenerate
semiconductors and semimetals placed in high external electric, and 
longitudinal and transverse magnetic fields are investigated by taking 
into account the heating of electrons and phonons as well as the thermal 
and mutual drags of charge carriers and phonons. The spectrum of charge 
carriers is assumed to be parabolic, e.g., ${\ds \varepsilon=p^2/2m}$.
The consideration is made for both deformation ($d-$) and piezoelectric
($p-$) interaction potentials of electrons with phonons.

The organization of the paper is as follows: The system of equations
of the problem and their solutions are given in Sec. II, the energy
balance equations and their solutions for different scattering
mechanisms are investigated in detail in Sec. III, the thermopower in
longitudinal magnetic field is presented in Sec. IV. The Sec. V
concentrates on the thermopower and NE effect in transverse magnetic
field. Finally, the conclusion is given in Sec. VI.

\section{Theory}
Consider a degenerate semiconductor or semimetal with fully ionized
impurities placed in high electric and nonquantizing magnetic fields.
We assume that there are temperature gradients of both electrons
(${\ds \nabla T_e}$), and long wavelength (LW) phonons interacting
with electrons (${\ds \nabla T_{ph}}$). The gradients may be realized
by the gradient of heating electric field ($\nabla E$); for example,
by placing one end of the specimen to the wave guide with heating
electromagnetic wave, or by producing lattice temperature gradient
($\nabla T$).

If the frequency of interelectronic collisions ${\ds \nu_{ee}}$ is
much bigger than that of electron-phonon collisions for the energy
transfer ${\ds \nu_{\varepsilon}}$, i.e., ${\ds \nu_{ee} \gg
\nu_{\varepsilon}}$, then the isotropic part of the distribution
function of electrons has the form of Fermi one with effective
temperature of electrons $T_e$,
\beq
f_0(\varepsilon)=\left[1+\exp\left(\frac{\zeta(T_e)
-\varepsilon}{T_e}\right)\right]^{-1},
\eeq  
where $\zeta(T_e)$ is the chemical potential and $\varepsilon$ is
the energy of charge carriers. Note that $T_e$ is in energy units.

We assume that in the lattice there is a thermal reservoir of
short wavelength (SW) phonons for LW phonons interacting with
electrons\cite{Gur}. The maximum momentum of LW phonons interacting
with electrons satisfies the condition: ${\ds q_{max} \approx 2p_0<T/s
\equiv q_T}$, where $T$is the lattice (reservoir) temperature, $q_T$
is the momentum of thermal phonons, ${\ds p_0}$ is the momentum of
electrons in the Fermi level, and $s$ is the velocity of sound in the
crystal. As it is shown in Ref. 45, under these conditions LW phonons
are heated as well. Therefore, we assume that the isotropic part of
the distribution function of phonons has the form:
\beq
N_0(q)=\left[\exp\left(\frac{\hbar\omega_q}{T_{ph}}\right)-1
\right]^{-1} \approx \frac{T_{ph}}{\hbar\omega_q}.
\eeq 
In accordance with Ref. 45, the distribution function of phonons
has the form of Eq. (2) only in two cases. In the first case the
frequency of LW phonon-electron collisions $\beta_e$ is much
smaller than the frequency of LW phonon-SW phonon collisions
$\beta_{ph}$. In this case $T_{ph}=T$ if
\beq
\frac{N(T_e)}{N(T)}\frac{\beta_e}{\beta_{ph}}\approx
\frac{T_e}{T}\frac{\beta_e}{\beta_{ph}} \ll 1.
\eeq
In the second case $\beta_e \gg \beta_{ph},\beta_b^{(\varepsilon)}$,
where ${\ds \beta_b^{(\varepsilon)}}$ is the collision frequency of
phonons with crystal boundaries connected with energy transfer to
outside. In this case, the temperature of LW phonons becomes equal
to the temperature of electrons ($T_{ph}=T_e$), and LW phonons are
in nonequilibrium state.

In high external fields electrons and phonons are essentially in
a nonequilibrium and anisotropic state. Therefore, the distribution
function of electrons $f(\varepsilon)$ and that of phonons
$N({\bf q})$ are, as usual, in the form
\beq
f(\varepsilon)=f_0(\varepsilon)+\frac{{\bf f_1}(\varepsilon).
{\bf p}}{p},~~~~~~N({\bf q})=N_0(q)+\frac{{\bf N_1}(q).{\bf q}}{q},
\eeq 
where ${\bf f_1}(\varepsilon)$ and ${\bf N_1}(q)$ are the
antisymmetric parts of the distribution functions of electrons and
phonons, respectively.

In the present paper, we assume that the so-called ``diffusion 
approximation" for electrons and phonons applies. Therefore, 
$\mid {\bf f_1}(\varepsilon) \mid \ll f_0(\varepsilon)$ and 
$\mid {\bf N_1}(q)\mid \ll N_0(q)$. The isotropic and anisotropic
parts of the distribution functions of electrons and phonons are
obtained from the coupled system of Boltzmann equations, which
form the main equations of the problem:
\bea
\frac{p}{3m}[\nabla {\bf f_1}(\varepsilon)]-\frac{2e}{3p}
~\frac{\partial}{\partial \varepsilon}
[\varepsilon {\bf E}.{\bf f_1}(\varepsilon)]= \\
\nonumber
\frac{m}{2\pi^2\hbar^3 p}~\frac{\partial}{\partial \varepsilon}
\left[\int_0^{2p} dq~\hbar\omega_q~W_q~q
\left\{\hbar\omega_q N_0(q)\left(\frac{\partial f_0(\varepsilon)}
{\partial \varepsilon}\right)+f_0(\varepsilon)[1-f_0(\varepsilon)]
\right\}\right],
\eea
\bea
\frac{p}{m}[\nabla f_0(\varepsilon)]-\frac{e{\bf E}p}{m}
\left(\frac{\partial f_0(\varepsilon)}{\partial\varepsilon}\right)
+\nu(\varepsilon){\bf f_1}(\varepsilon)-\Omega
[{\bf h}.{\bf f_1}(\varepsilon)] =\\
\nonumber
-\frac{4\pi}{(2\pi\hbar)^3}\frac{1}{p^2}\left(\frac{\partial
f_0(\varepsilon)}{\partial \varepsilon}\right)~\int_0^{2p}~dq
~W_q~q^2~\hbar\omega_q~ {\bf N_1}(q),
\eea
\beq
s\nabla N_0(q)+\beta(q){\bf N_1}(q)=\frac{4\pi m W_q N_0(q)}{(2\pi
\hbar)^3}\int_{\varepsilon(q/2)}^{\infty}~ dp~{\bf f_1}(\varepsilon),
\eeq
\beq
\frac{s}{3}\nabla{\bf N_1}(q)+\beta_0(q)N_0(q)-\left[(\beta_{ph}+
\beta_b^{(\varepsilon)})N(q,T)+\beta_e N(q,T_e)\right]=0.
\eeq
In Eqs. (5)-(8), $e$ is the absolute value of the electronic charge, 
$m$ is the effective mass of electrons, $\hbar\omega_q$ and $q$ are
the energy and the quasimomentum of phonons, respectively, $N(q,T)$
and $N(q,T_e)$ are the equilibrium Planck distribution functions
with temperatures $T$ and $T_e$. $W_q$ is the quantity from which
the scattering probability of electrons by acoustical phonons is
obtained. 
It is defined as
\beq
W_q=W_0~q^t.
\eeq
For $d-$ interaction $t=1$ and ${\ds W_0 \approx \frac{2\pi G^2}
{\rho~s~\hbar}}$, where $G$ is the deformation potential constant.
On the other hand, for $p-$ interaction $t=-1$ and ${\ds W_0=
\frac{(4\pi)^3\hbar e^2\sum^2}{\rho~s~\epsilon_0^2}}$, where
$\sum$, $\rho$ and $\epsilon_0$ are the piezoelectric module,
density and the dielectric constant of the crystal, respectively. 

The collision frequency of electrons with phonons
${\ds\nu_{ph}(\varepsilon)}$ is
\beq
\nu_{ph}(\varepsilon)=\frac{m W_0}{2\pi^2 \hbar^3 p^3}
\int_0^{2p}~dq~q^{3+t}\left[N_0(q)+\frac{1}{2}\right].
\eeq
The total collision frequency of phonons is defined as
\beq
\beta(q)=\beta_0(q)+\beta_b(q)=\beta_e(q)+\beta_{ph}(q)+
\beta_b^{(\varepsilon)}(q)+\beta_b^{(p)}(q),
\eeq
where the indices mean the collision frequency of phonons with
electrons ($e$), phonons ($ph$), and crystal boundaries ($b$) for
the energy or momentum transfer to outside. The total collision
frequencies of phonons $\beta(q)$ and electrons $\nu(\varepsilon)$
may be given in the form
\beq
\beta(q)=\beta(T)\left(\frac{q s}{T}\right)^k, ~~~~~~~~
\nu(\varepsilon)=\tilde{\nu}\left(\frac{\varepsilon}{\zeta_0}
\right)^r,~~~~~~~~\tilde{\nu}=\nu_0 \Theta_{e,ph}^{\ell},
\eeq
where $\zeta_0=\varepsilon_F$ is the Fermi energy,
$\nu_0=\nu(\zeta_0)$ and $\Theta_{e,ph}=T_{e,ph}/T$ is the
dimensionless temperature of electrons and phonons. For the
scattering of electrons by the impurity ions $r=-3/2$, $\ell=0$; by
the deformation potential of acoustical phonons (d-interaction)
$r=1/2$, $\ell=1$; by the piezoelectric potential of acoustical
phonons (p-interaction) $r=-1/2$, $\ell=1$; and $k=0,1,t$ for
scattering of LW phonons by crystal boundaries, by SW phonons and
electrons, respectively. 

By neglecting the first term in Eq. (8), we obtain Eq. (2). By
using Eqs. (1) and (6), for ${\bf f_1}(\varepsilon)$ we have
\bea
{\bf f_1}(\varepsilon)-\frac{\Omega}{\nu(\varepsilon)}
[{\bf h}.{\bf f_1}(\varepsilon)]-\frac{p}{m \nu(\varepsilon)}
\left(e{\bf E^{\prime}}+\left[\frac{\varepsilon-\zeta}
{T_e}\right]\nabla T_e\right)\frac{\partial f_0(\varepsilon)}
{\partial\varepsilon}=~~~\\
\nonumber
\frac{m^2}{2\pi^3\hbar^3}\frac{1}{p^3(\varepsilon)}
\left(\frac{\partial f_0(\varepsilon)}{\partial
\varepsilon}\right)\int_0^{2p}~dq~W_q~\hbar\omega_q~\frac{q^2}
{\beta(q)}\left\{s\nabla N_0(q)-\frac{m W_q N_0(q)}
{2\pi^2 \hbar^3}\int_{\varepsilon(q/2)}^{\infty}~dp~{\bf f_1}
(\varepsilon)\right\}.
\eea
The first term on the right hand side of Eq. (13) is in accord
with the thermal drag and the second term with the mutual drag.
Eq. (13) is the integral equation for ${\bf f_1}(\varepsilon)$,
but if we assume as usual

\beq
{\bf f_1}(\varepsilon)=p {\bf V}(\varepsilon)\left[-
\frac{\partial f_0(\varepsilon)}{\partial \varepsilon}\right],
\eeq
then for the case of degenerate electrons, this equation becomes
an algebraic one. In Eq. (14), ${\bf V}(\varepsilon)$ is the drift
velocity of electrons. Replacing the integral
${\ds\int_{\varepsilon(q/2)}^{\infty} d\varepsilon~V(\varepsilon)
\left(-\frac{\partial f_0(\varepsilon)}
{\partial \varepsilon}\right)}$ by $V(\zeta_0)$, we obtain the
following equation for ${\bf V}(\varepsilon)$:
\bea
{\bf V}(\varepsilon)=-\frac{\nu(\varepsilon)}{m[\Omega^2+\nu^2
(\varepsilon)]}~~~~~\\
\nonumber
\left\{\left[{\bf F}+\frac{\Omega}{\nu(\varepsilon)}[{\bf h}.{\bf F}]
+{\bf h}.[{\bf h}.{\bf F}]\frac{\Omega^2}{\nu^2(\varepsilon)}\right]
-m\nu(\varepsilon)\gamma(\varepsilon)\left[{\bf V_0}+\frac{\Omega}
{\nu(\varepsilon)}[{\bf h}.{\bf V_0}]+\frac{\Omega^2}
{\nu^2(\varepsilon)}{\bf h}.({\bf h}.{\bf V_0})\right]\right\},
\eea
where
\beq
{\bf F}=e{\bf E^{\prime}}+\left(\frac{\varepsilon-\zeta}{T_e}\right)
\nabla T_e + A_{kt}\left(\frac{\varepsilon}{T}\right)^{(t-k)/2}
\nabla T_{ph},
\eeq
\beq
A_{kt}=\frac{2^{3+3(t-k)/2}}{3+t-k}~\frac{\beta_e(T)}{\beta(T)}
\left(\frac{m s^2}{T}\right)^{(t-k)/2},
\eeq
\beq
{\bf E^{\prime}}={\bf E}+{\bf E_T}+\frac{1}{e}\nabla\zeta(T_e),
\eeq
where ${\bf E}$ is the external field, ${\bf E_T}$ is the
thermoelectric field, and $V_0=V(\zeta_0)$.

The expression characterizing the mutual drag of charge carriers
and phonons is
\beq
\gamma(\varepsilon)=\frac{3+t}{2^{3+t}}~\frac{\nu_{ph}(\varepsilon)}
{\nu(\varepsilon)}~\frac{1}{p^{3+t}}\int_0^{2p}~dq~\frac{\beta_e(q)}
{\beta(q)}~q^{2+t}.
\eeq
The mutual drag is strong as $\gamma \rightarrow 1$. This means
that electrons and phonons are scattered preferably by each other,
i.e., $\nu(\varepsilon) \approx \nu_{ph}(\varepsilon)$ and
$\beta(q)\approx \beta_e(q)$. In fact, there are other scattering
mechanisms of electrons and phonons. Because, in the present work
we assume the diffusion approximation $\gamma(\varepsilon)$ must
be smaller than 1. 
 
To obtain ${\bf V}(\zeta) \equiv {\bf V_0}(\zeta)$ from Eq. (13)
by the accuracy of the second approximation on degeneracy, we get
the following relation for the electrical current:
\bea
{\bf J}=\sigma_{11}{\bf E^{\prime}}+\sigma_{12}
[{\bf h}.{\bf E^{\prime}}]+\sigma_{13}{\bf h}.
[{\bf h}.{\bf E^{\prime}}]+\beta_{11}^{(e)}\nabla T_e
+\beta_{12}^{(e)}[{\bf h}\nabla T_e]+\\
\nonumber
\beta_{13}^{(e)}{\bf h}.[{\bf h}\nabla T_e]+\beta_{11}^{(ph)}
\nabla T_{ph}+\beta_{12}^{(ph)}[{\bf h}\nabla
T_{ph}]+\beta_{13}^{(ph)}{\bf h}.[{\bf h}\nabla T_{ph}],
\eea
where 
\bea
\sigma_{1i}=\int_0^{\infty}~d\varepsilon~a(\varepsilon)
\left[1+b(\varepsilon) g_i(\varepsilon)\right], \\
\nonumber
\beta_{1i}^{(e)}=\frac{1}{e}\int_0^{\infty}~d\varepsilon
~a(\varepsilon)\left[\frac{\varepsilon-\zeta}{T_e}+
g_i(\varepsilon)d_0(\varepsilon)\right],\\
\nonumber
\beta_{1i}^{(ph)}=\frac{A_{kt}}{e}\int_0^{\infty}~d\varepsilon
~a(\varepsilon)\left\{\left(\frac{\varepsilon}{T}
\right)^{(t-k)/2}+\left(\frac{\zeta_0}{T}\right)^{(t-k)/2}
b(\varepsilon)g_i(\varepsilon)\right\},
\eea
and
\bea
a(\varepsilon)=\frac{2^{3/2}~m^{1/2}~e^2}{3~\pi^2~\hbar^3}
\left(\frac{\Omega}{\nu(\varepsilon)}\right)^{i-1}
\frac{\varepsilon^{3/2}~\nu(\varepsilon)}{\Omega^2+
\nu^2(\varepsilon)}~\left(-\frac{\partial f_0(\varepsilon)}
{\partial \varepsilon}\right),\\
\nonumber
d_0(\varepsilon)=\frac{\pi^2}{12}\frac{T_e}{\zeta_0}
b(\varepsilon),~~~~b(\varepsilon)=\frac{\gamma(\varepsilon)
\nu(\varepsilon)}{\Omega^2+\tilde{\nu}^2(\varepsilon)
(1-\gamma_0)^2},\\
\nonumber
g_1(\varepsilon)=\tilde{\nu}(1-\gamma_0)-\frac{\Omega^2}
{\nu(\varepsilon)},~~~~g_2(\varepsilon)=\nu(\varepsilon)+
\tilde{\nu}(1-\gamma_0),\\
\nonumber
g_3(\varepsilon)=\nu(\varepsilon)\left[1+
\frac{\tilde{\nu}(1-\gamma_0)}{\nu(\varepsilon)}+
\frac{\Omega^2+\nu^2(\varepsilon)}{\nu(\varepsilon)
\tilde{\nu}(1-\gamma_0)}\right],~~~~
\tilde{\nu}=\nu(\zeta_0,\Theta_{ph}),
~~~~\gamma_0=\gamma(\zeta_0)\Theta_{ph}^{1-\ell}.
\eea

\section{The energy balance equations and their solutions}
To define the total thermopower and thermomagnetic effects
as a function of $E$, $H$, and $\gamma_0$, we must start
from the energy balance equation obtained in Ref. 45. We
consider two different cases:\\
{\bf (a)}~~LW phonons are not heated and electrons transfer
their energy gained from the external field to the reservoir
of SW phonons, which has the equilibrium state at the lattice
temperature $T$. Then, the energy balance equation has the form:
\beq
\frac{(eE)^2~\nu_0(1-\gamma_0)}{\Omega^2+\nu_0^2(1-\gamma_0)^2}
=\frac{3~2^{2+t}}{(3+t)\pi^2\hbar^3}~m^3~s~T~p_0^t~W_0[\Theta_e-1],
\eeq
where $\gamma_0=\gamma(\zeta_0)$.\\
{\bf (b)}~~LW phonons are heated and they transfer their energy
gained from electrons to the reservoir of SW phonons,
\beq
\frac{(eE)^2\tilde{\nu}(1-\gamma_0)}{\Omega^2+\tilde{\nu}^2
(1-\gamma_0)^2}=6~m~s~p_0~\beta_{ph}(T)[\Theta_e-1],
\eeq
in this case $\gamma_0=\gamma(\zeta,\Theta_{ph})$.

We consider now the dependences of $\Theta_e$ on $E$, $H$ and
$T$, which are obtained by solving Eqs. (23) and (24). The
solution of Eq. (23) for the arbitrary scattering mechanisms,
degree of electron heating and the ratio ${\ds \Omega/\nu_0}$ is
\beq
\Theta_e=1+\left(\frac{E}{E_i}\right)^2\frac{(1-\gamma_0)}
{1+\left(\nu_0/\Omega\right)^2(1-\gamma_0)^2},
\eeq
where ${\ds E_i=(3~2^{2+t}~m^3~s~T~p_0^t~W_0~\Omega^2/(3+t)
~\pi^2~\hbar^3~e^2~\nu_0)^{1/2}}$. We may consider $\Theta_e$
in two limits: $\Omega \gg \nu_0$ and $\Omega \ll \nu_0$.
In the first limit
\beq
\Theta_e=1+\left(\frac{E}{E_i}\right)^2\left(1-\gamma_0\right),
\eeq
in the second limit
\beq
\Theta_e=1+\left(\frac{E}{E_j}\right)^2 \left(1-\gamma_0\right)^{-1},
\eeq
where ${\ds E_j=E_i \nu_0/\Omega}$.

From Eq. (26), for different scattering mechanisms $E_i$ take
the forms:\\
\\
for~ $r=1/2$, $r=-1/2$:~~~~${\ds E_1^2=3\left(\frac{H~s}{c}\right)^2}$,\\
\\
for~ CI/DA:~~${\ds E_2^2=\frac{9~\pi}{2}\left(\frac{\varepsilon_0~G~H}
{c~e^2}\right)^2~\frac{T~n^{1/3}}{F~\rho}}$,\\
\\
for~ CI/PA:~~~${\ds E_3^2=3\left(\frac{4~\pi~\Sigma~H}{c~e}\right)^2~
\frac{T}{F~\rho~n^{1/3}}}$.\\
\\
Hereafter, CI/DA (CI/PA) means that energy of electrons is scattered
by the deformation acoustical, (piezo acoustical) phonons and
momentum of electrons by the charged impurity ions (CI). Similarly,
from Eq. (27):\\
\\
for~ DA:~~${\ds E_4^2=3\left(\frac{m~G^2~T~n^{1/3}}{\rho~e~s\hbar^3}
\right)^2}$,\\
\\
for~ PA:~~${\ds E_5^2=\frac{1}{3}\left(\frac{32~\pi~m^2~e~\Sigma^2~T}
{\varepsilon_0^2~\hbar^3~\rho~ s~n^{1/3}}\right)^2}$,\\
\\
for~ CI/DA:~~${\ds E_6^2=\frac{2}{3}\left(\frac{e~m^2~G}
{\varepsilon_0~\hbar^3}\right)^2~\frac{F~T~n^{1/3}}{\rho}}$,\\
\\
for~ CI/PA:~~${\ds E_7^2=\frac{1}{3}\left(\frac{8~e^2~m^2~\Sigma^2}
{\varepsilon_0^2~\hbar^3}\right)^2~\frac{F~T}{\rho~n^{1/3}}}$.\\
\\
In the case of phonon heating ($T_{ph}=T_e$) if $E \perp H$ and 
$\Omega\gg\nu$, then for DA and PA scattering mechamisms of
electrons by phonons one finds $\Theta_e$ as
\beq
\Theta_e=\left[1-\left(\frac{E}{E_i}\right)^2\left(1-\gamma_0\right)
\right]^{-1},
\eeq 
where the characteristic fields $E_i$ are: \\
\\
for~ DA:~~${\ds E_8^2=\frac{3}{2}\left(\frac{H~T}{c~s~m~G}\right)^2}$,\\
\\
for~ PA:~~${\ds E_9^2=3\left(\frac{\varepsilon_0~H~T^2~n^{1/3}}
{8~c~s~m~e~\Sigma}\right)^2\rho}$.\\
\\
If the momentum of electrons are scattered from the impurity ions,
regardless of the scattering of energy from either DA or PA phonons, 
we find
\beq
\Theta_e=\frac{1+\left(E/E_{10}\right)^2}{1+\gamma_0\left(E/E_{10}
\right)^2},~~~~~E_{10}^2=\frac{9\pi}{4}\left(\frac{\varepsilon_0~H~
T^2}{c~s^2~m~e^2}\right)^2\frac{T~n^{1/3}}{F~\rho}.
\eeq
As it is seen from Eq. (29), under the condition of mutual drag
the electron temperature is finite, i.e. because
${\ds \gamma_0(E/E_{10})^2 \gg 1}$, and
${\ds \Theta_e \leq 1/\gamma_0}=const$.

If $E \perp H$ and $\Omega \ll \nu$, for DA and PA scattering 
mechanisms $\Theta_e$ becomes
\beq
\Theta_e=\frac{1}{2}\left\{1+\left[1+4\left(\frac{E}{E_i}\right)^2
\left(1-\gamma_0\right)^{-1}\right]^{1/2}\right\},
\eeq
where the characteristic fields $E_i$ are:\\
\\
for~ DA:~~${\ds E_{11}^2=\frac{3}{2}\left(\frac{m~G~T^3~n^{1/3}}
{e~s^3~\hbar^3~\rho}\right)^2}$,\\
\\
for~ PA:~~${\ds E_{12}^2=3\left(\frac{4~m~\Sigma~T^3}
{\varepsilon_0~s^3~\hbar^3~\rho}\right)^2}$.\\

For both CI/DA and CI/PA scattering mechanisms the critical field
is the same; and $\Theta_e$ is found to be as
\bea
\Theta_e=\frac{1}{2\gamma_0}\left\{(1+\gamma_0)-
\left[(1-\gamma_0)^2-4\gamma_0\left(\frac{E}{E_{13}}\right)^2
\right]^{1/2}\right\},\\
\nnb
E_{13}^2=\frac{1}{\pi}\left(\frac{e~m~T^2}{\varepsilon_0
~\hbar^3~s^2}\right)^2~\frac{F~T~n^{1/3}}{\rho}.
\eea
Finally, in the absence of mutual drag ($\gamma_0 \rightarrow 0$),
Eq. (31) gives
\beq
\Theta_e=\left(\frac{E}{E_{13}}\right)^2.
\eeq

\section{Thermopower in longitudinal magnetic field}
We will first consider the case ${\bf E}\perp{\bf H}\parallel
\nabla T_{e,ph} \parallel {\bf \hat{z}}$. From $J_z=0$ condition,
we have
\beq
E_{Tz}+\frac{1}{e}\nabla_z \zeta(T_e)=\alpha_e\nabla_z T_e+
\alpha_{ph}\nabla_z T_{ph},~~~~~~\alpha_{e,ph}=
\frac{\beta_{11}^{(e,ph)}+
\beta_{13}^{(e,ph)}}{\sigma_{11}+\sigma_{13}},
\eeq
where $\alpha_e$ and $\alpha_{ph}$ are the electron and phonon
parts of the differential thermopower, respectively. By taking
into account the fact that ${\ds
\gamma(\varepsilon)=\gamma_0(\varepsilon/\zeta_0)^{t-k/2-r}}$,
we find
\beq
\alpha_e=-\frac{1}{e}\frac{\pi^2}{6}~
\left[3-2r-\gamma_0\left(\frac{5}{2}-r\right)\right]~\frac{T}
{\zeta_0}~\Theta_e,~~~~~~\alpha_{ph}=-\frac{1}{e}~A_{kt}
\left(\frac{\zeta_0}{T}\right)^{(t-k)/2}.
\eeq 
The thermopower is given by
\beq
V=\int_0^{L_z}dz~(\alpha_e \nabla_z T_e+\alpha_{ph}\nabla_z T_{ph})
=V_e+V_{ph},   
\eeq
where $L_z$ is the size of the specimen in the $z$ direction. As it
follows from Eqs. (34) and (35) in weak longitudinal magnetic field
${\ds (\Omega \ll \nu_0[1-\gamma_0])}$, in the absence of phonon
heating the electronic part of the total thermopower $V_e$ is
proportional to ${\ds E_0^4/(1-\gamma_0)^2}$; and the phonon part
$V_{ph}$, in general, does not depend on $\gamma_0$. 

At high magnetic field ${\ds (\Omega \gg \nu_0[1-\gamma_0])}$,
$V_e$ is proportional to ${\ds (E_0/H)^4(1-\gamma_0)^2}$, with $E_0$
being the heating electric field intensity at the end of the specimen
where electrons are highly heated. Therefore, with increasing
$\gamma_0$, at weak magnetic field $V_e$ grows as $\sim {\ds
(1-\gamma_0)^{-2}}$, and at high magnetic field $V_e$ decreases as
$\sim{\ds (1-\gamma_0)^2}$. 

In the case of strong heating of LW phonons and for the scattering
of momentum and energy of electrons by acoustical phonons, at weak
magnetic fields, from Eqs. (24), (34) and (35), we have
\bea
\nonumber 
V_{ph} \sim \Theta_e \sim \frac{E_0}{1-\gamma_0},~~~~~~
V_e =\left(\frac{E_0}{1-\gamma_0} \right)^2, 
\eea
and at high magnetic fields
\bea
\nonumber
V_{ph} \sim \left[1-\frac{E_0^2}{E_{01}^2}(1-\gamma_0)\right]^{-1},
~~~~~~V_e \sim \left[1-\frac{E_0^2}{E_{01}^2}
(1-\gamma_0)\right]^{-2},\\
\nonumber 
E_0^2(1-\gamma_0)<E_{01}^2=\frac{6 s~p_0~\beta_{ph}~H^2}
{(m~c^2~\nu_0)^2}. 
\eea

In the calculation for the dependences of $V_e$ and the NE voltage
($U$) on $E$ and $H$ in the transverse magnetic fields, it is
necessary to assume that $\nabla T_{e,ph}$ is constant along the
specimen, i.e., at one end of the specimen electrons are heated
strongly by the electric field $(\Theta_e \gg 1)$, however, at
the other end their temperature is $T$.

\section{Thermopower and Nernst-Ettingshausen effect in
transverse magnetic field}
In general, the thermomagnetic effects are measured experimentally
under the condition of ${\ds \nabla_x T_{e,ph}=0}$. We will direct
the external fields ${\bf E}$ and ${\bf H}$ along the y-axis and
the temperature gradients along the z-axis. Therefore, from Eq. (20)
and the condition ${\ds J_x=J_z=0}$, for the transverse NE voltage,
we obtain
\beq
E_{Tx}=-H(Q_e\nabla_z T_e+Q_{ph}\nabla_z T_{ph}),~~~~~~
Q_{e,ph}=\frac{1}{H}~\frac{\sigma_{11}\beta_{12}^{(e,ph)}-
\sigma_{12}\beta_{11}^{(e,ph)}}{\sigma_{11}^2+\sigma_{12}^2}.
\eeq
In this case the thermoelectric field $E_{Tz}$ coincides with
Eq. (33) by changing $\alpha_{e,ph}$ as
\beq
\alpha_{e,ph}=-\frac{\sigma_{11}\beta_{11}^{(e,ph)}+\sigma_{12}
\beta_{12}^{(e,ph)}}{\sigma_{11}^2+\sigma_{12}^2}.
\eeq

We would like to investigate Eqs. (35) and (36) by taking into
account Eqs. (21) and (22) in the weak and high magnetic field
limits in the following subsections.

\subsection*{\bf The weak magnetic field case}
If ${\ds \tilde{\nu}^2 \gg \Omega^2}$, then for the electron
part $Q_e$, and phonon part $Q_{ph}$ of the NE coefficients,
we obtain
\bea
Q_e=-\frac{1}{e}\frac{\pi^2}{3}\frac{\mu_0}{c}\frac{T}{\zeta_0}
\left\{r+\gamma_0\left(\frac{5}{4}-2r\right)\right\}
\Theta_e~\Theta_{ph}^{-\ell},\\
\nonumber
Q_{ph}=-\frac{1}{e}\frac{\pi^2}{6}\frac{\mu_0}{c}~(t-k)
\left(\frac{T}{\zeta_0}\right)^{2+(k-t)/2}\left\{r+\gamma_0
\left(1-2r+\frac{t-k}{4}\right)\right\}~A_{kt}~
\Theta_e^2~\Theta_{ph}^{-\ell},
\eea
where ${\ds \mu_0=e/m\nu_0}$ is the mobility of ``cold" electrons.
As it is seen from this equation, under the conditions of strong
mutual drag for the parabolic spectrum of electrons ($\beta_e \gg
\beta_{ph}, \beta_{pb}$, i.e., $\beta_e \approx \beta$), the
phonon part of the NE coefficient $Q_{ph}=0$, or more exactly
${\ds Q_{ph}\sim (\beta-\beta_e)/\beta=(\beta_{ph}+\beta_{pb})/
\beta \ll 1}$ (see also Refs. 20 and 21). Moreover, electrons and
phonons form a system coupled by the mutual drag with common
temperature $T_e=T_{ph}$ and drift velocity ${\ds v_e=v_{ph}=s}$.
For this reason, there is only one thermomagnetic coefficient for
the quasiparticle (electron dressed by phonon) coupled by the
mutual drag. The quasiparticle has the electronic charge $e$, and
the mass of phonons ${\ds M=T_e/s^2}$ (see Refs. 27-30). However,
since we assume the diffusion approximation, $\gamma_0 < 1$ or
$u<s$, $Q_{ph}$ is proportional to ${\ds (\beta_{ph}+\beta_{pb})/
\beta \neq 0}$. Only when $\gamma_0=1$ or $u=s$, we have $Q_{ph}=0$.

The expression of ${\ds \alpha_{e,ph}}$ at weak magnetic field
coincides with Eq. (34). Therefore, here we give only the
expressions denoting the change in ${\ds \alpha_e}$ and
${\ds \alpha_{ph}}$ in the weak magnetic field:
\bea
\Delta\alpha_e=-\frac{1}{e}\frac{\pi^2}{3}\left\{\frac{7}{4}-2r
+\left(\frac{1}{4}-r\right)\frac{\gamma_0(2-\gamma_0)}
{(1-\gamma_0)^2}\right\}\frac{\Omega^2}{\nu_0^2}~\frac{T}{\zeta_0}
\Theta_e~\Theta_{ph}^{-2\ell},\\
\nonumber
\Delta\alpha_{ph}=-\frac{1}{e}~A_{kt}\left(\frac{\zeta_0}{T}
\right)^{(t-k)/2}\frac{\Omega^2}{\nu_0^2(1-\gamma_0)^2}
\Theta_{ph}^{-2\ell}.
\eea
In the case of scattering of electrons by deformation acoustical
phonons ($r=1/2$) as $\gamma_0 \rightarrow 1$, the last term in
square bracket in Eq. (39) is negative and much bigger than the
other terms, hence, $\Delta\alpha_e$ changes its sign. The NE
voltage has the form:
\beq
U=-\int_0^{L_x}dx~H(Q_e\nabla_z T_e+Q_{ph} \nabla_z T_{ph})
=U_e+U_{ph}.
\eeq	

In the cases of the absence and presence of phonon heating, the
energy balance equation in the transverse magnetic field has the
form, respectively:
\beq
E^2=E_{02}^2(1-\gamma_0)(\Theta_e -1),~~~~~~E_{02}^2=
\frac{3~2^{2+t}~m^3~s~\nu_0~p_0^t~T~W_0}{(3+t)\pi~\hbar^3~e^2},
\eeq
\beq
(eE)^2=6\beta_{ph}(T)~m~s~p_0~\tilde{\nu}(1-\gamma_0)(\Theta_e-1).
\eeq
It follows from Eq. (38) that $Q_e$ has two components. Then,
by using Eqs. (38), (40), and (41) we obtain the first and the
second components of the electron part of NE voltage as
\beq
U_e^I \sim (1-\gamma_0)^{-2}E_0^4,~~~~~~U_e^{II} \sim 
\gamma_0(1-\gamma_0)^{-2}E_0^4.
\eeq
By analogy, we may obtain the phonon part as
\beq
U^{ph} \sim (1-\gamma_0)^{-2}E_0^4.
\eeq
It is interesting that in the absence of phonon heating both
$U_e$ and $U_{ph}$ are proportional to ${\ds H E_0^4}$, i.e.,
they have the same dependence on the intensity of electric
and magnetic fields for all scattering mechanisms of electrons
in the case of strong electron heating, ${\ds \Theta_e \gg 1}$.
If the energy and momentum of electrons are transferred to
phonons, in the case of strong electron and phonon heating,
we have
\beq
U_e^I \sim E_0 H (1-\gamma_0)^{-1/2}, ~~~~~~
U_e^{II} \sim E_0 H \gamma_0(1-\gamma_0)^{1/2},
\eeq
and
\beq
U_{ph}^I \sim E_0^2 H(1-\gamma_0),~~~~~~
U_{ph}^{II} \sim E_0^2 H \gamma_0(1-\gamma_0)^{-1}.
\eeq
Therefore, the mutual drag of electrons and phonons causes
essential changes in the thermomagnetic behavior of semiconductors
and semimetals. The strong phonon heating leads to the important
contribution to these effects, because in this case $\gamma_0 \sim
\Theta_e$ for the scattering of momentum of electrons by impurity
ions and energy by LW phonons (the thermal drag case). 	

\subsection*{\bf The high magnetic field case:}
In the limit ${\ds \Omega^2 \gg \tilde{\nu}^2}$, the thermomagnetic 
coefficients take the forms:
\beq
\alpha_e=-\frac{1}{e}\frac{\pi^2}{2}\frac{T}{\zeta_0}~\Theta_e,
\eeq
\bea
Q_e=-\frac{1}{e}\frac{\pi^2}{3}\left(r+\frac{5}{4}\gamma_0\right)
\frac{c}{H^2 \mu_0}\frac{T}{\zeta_0}~\Theta_e~\Theta_{ph}^{\ell},\\
\nonumber
Q_{ph}=-\frac{1}{e}\frac{\pi^2}{6}(t-k)\left\{r+\gamma_0
\left(1+\frac{t-k}{4}\right)\right\}\frac{c~A_{kt}}{H^2\mu_0}
\left(\frac{T}{\zeta_0}\right)^{2+(k-t)/2}\Theta_e^2
~\Theta_{ph}^{\ell}.
\eea
As it is seen from Eq. (48), for the case of weak mutual drag,
both the electronic and phonon parts of the transverse NE
coefficients change their sign for the scattering of electrons by
the piezo acoustical phonons ($r=-1/2$). Moreover, the phonon part
of NE coefficient changes its sign if LW phonons are scattered by
SW phonons $(k=1)$, and electrons are scattered by piezo
acoustical phonons $(t=-1)$.  

For the case under consideration, the expression for $\alpha_{ph}$ 
coincides with Eq. (34). As it follows from Eqs. (38) and (48), at
weak and high magnetic fields $Q_{ph}=0$ for the scattering of LW
phonons by electrons ($k=t$), and by SW phonons $(t=1)$. From
Eqs. (40)-(42) and (48), in the absence of phonon heating, we obtain, 
\beq
U_{e,ph}^I \sim H^{-1}E_0^4(1-\gamma_0)^{-2},~~~~~~
U_{e,ph}^{II}\sim H^{-1}E_0^4 \gamma_0(1-\gamma_0)^{-2}.
\eeq 
In the case of strong phonon heating, for the scattering of energy
and momentum of electrons by phonons, we find
\beq
U_e \sim H^{-1}E_0^3(1-\gamma_0)^{-3/2},~~~~~~
U_{ph} \sim H^{-1}E_0^4(1-\gamma_0)^{-2}.
\eeq

\section{Conclusion}
In the present work, we have shown that at weak longitudinal
magnetic fields in the absence of phonon heating, the electron part
of thermoelectric power $V_e$ increases with increasing $E_0$ and
the degree of mutual drag of electrons and phonons $\gamma_0$.
Nevertheless, the phonon part $V_{ph}$ does not depend on
$\gamma_0$. At longitudinal high magnetic fields, $V_e$ increases
with increasing $E_0$, and decreases with increasing $H$ and
$\gamma_0$. In the case of strong phonon heating, if the momentum
and energy of electrons are transferred to acoustical phonons at
weak magnetic fields, $V_e$ and $V_{ph}$ grow as $E_0$ and
$\gamma_0$ increase. It has been shown that at high magnetic field
for a given $\gamma_0 < 1$, $V_e$ and $V_{ph}$ grow as $H$ increases.

In weak transverse magnetic field, $V_e$ and $V_{ph}$ are exactly
the same as in the case of longitudinal magnetic field, and in the
absence of phonon heating both the electron and phonon parts of the
transverse NE voltage $U_e$ and $U_{ph}$ are proportional to
$HE_0^4$. In the case of strong electron and phonon heating both
$U_e$ and $U_{ph}$ grow as $E$, $H$ and $\gamma_0$ increase. At high
magnetic field in the absence of phonon heating, $U_e$ and $U_{ph}$
grow with increasing $E_0$ and $\gamma_0$, and decrease linearly
with increasing $H$. It has also been shown that in weak and high
transverse magnetic fields, both the electronic and phonon parts of
the NE coefficients change their sign for some scattering mechanisms.

\subsection*{Acknowledgments}
This work was partially supported by the Scientific and Technical 
Research Council of Turkey (TUBITAK). In the course of this work, 
T. M. Gassym was supported by TUBITAK-NATO.


\begin{thebibliography}{99}
\bibitem{1} C. W. J. Beenakker and A. A. M. Staring, Phys. Rev. B 
{\bf 46}, 9667 (1992).
\bibitem{2} L. W. Molenkamp, A. A. M. Staring, B. W. Alphenaar and
H. van Houten, {\it Proceedings of 8th International Conference on 
Hot Carriers in Semiconductors} (Oxford University Press, Oxford, 
1993).
\bibitem{3} M. J. Kearney and P. N. Butcher, J. Phys. C {\bf 19}, 
5429 (1986); {\bf 20}, 47 (1987).
\bibitem{4} R. J. Nicholas, J. Phys. C {\bf 18}, L695 (1985).
\bibitem{5} R. Fletcher, J. C. Maan, and G. Weimann, Phys. Rev. B
{\bf 32}, 8477 (1985).
\bibitem{6} R. Fletcher, J. C. Maan, K. Ploog, and G. Weimann,
Phys. Rev. B {\bf 33}, 7122 (1986).
\bibitem{7} D. G. Cantrell and P. N. Butcher, J. Phys. C {\bf 19},
L429 (1986); {\bf 20}, 1985 (1987); {\bf 20}, 1993 (1987).
\bibitem{8} L. D. Hicks and M. S. Dresselhaus, Phys. Rev. B
{\bf 47}, 12727 (1993).
\bibitem{9} X. Zianni, P. N. Butcher, and M. J. Kearney,
Phys. Rev. B {\bf 49}, 7520 (1994).
\bibitem{10} R. Fletcher, J. J. Harris, C. T. Foxon, M.
Tsaousidou, and P. N. Butcher, Phys. Rev. B {\bf 50}, 14991 (1994).
\bibitem{11} X. L. Lei, J. Phys.: Condens. Matter {\bf 6}, L305
(1994).
\bibitem{12} D. Y. Xing, M. Liu, J. M. Dong, and Z. D. Wang,
Phys. Rev. B {\bf 51}, 2193 (1995).
\bibitem{13} X. L. Lei, J. Cai, and L. M. Xie, Phys. Rev. B
{\bf 38}, 1529 (1988).
\bibitem{15} E. M. Conwell and J. Zucker, J. Appl. Phys. {\bf 36},
2192 (1995).
\bibitem{16} A. A. Abrikosov, {\it Introduction to the Theory
of Normal Metals: Solid State Physics Suppl.} (Academic,
New York, 1972), Vol. 12.
\bibitem{17} B. M. Askerov, {\it Electron Transport Phenomena
in Semiconductors}, (World Scientific, Singapore, 1994).
\bibitem{18} M. Bailyn, Phys. Rev. {\bf 112}, 1587 (1958);
{\bf 157}, 480 (1967).
\bibitem{19} L. E. Gurevich and T. M. Gassymov, Fiz. Tverd.
Tela (Leningrad) {\bf 9}, 3493 (1967).
\bibitem{20} M. M. Babaev and T. M. Gassymov, Phys. Status
Solidi B {\bf 84}, 473 (1977).
\bibitem{21} M. M. Babaev and T. M. Gassymov, Fiz. Technika
Poluprovodn. (Leningrad) {\bf 14}, 1227 (1980).
\bibitem{22} T. M. Gassymov, A. A. Katanov and  M. M. Babaev,
Phys. Status Solidi B {\bf 119}, 391 (1983).
\bibitem{23} M. M. Babaev, T. M. Gassymov and A. A. Katanov,
Phys. Status Solidi B {\bf 125}, 421 (1984).
\bibitem{Kane} M. M. Babaev, T. M. Gassym, M. Ta\0 and M.
Tomak, Phys. Rev. B {\bf 65}, 165324 (2002).
\bibitem{24} X. L. Lei, C. S. Ting, Phys. Rev. B {\bf 30},
4809 (1984); {\bf 32}, 1112 (1985).
\bibitem{25} T. H. Geballe and G. W. Hull, Phys. Rev.
{\bf 94}, 279 (1954); {\bf 94}, 283 (1954).
\bibitem{26} M. W. Wu, N. J. M. Horing and H. L. Cui,
cond-mat/9512114 (unpublished).
\bibitem{27} T. M. Gassymov, A. A. Katanov, J. Phys.:
Condens. Matter {\bf 2}, 1977 (1990).
\bibitem{28} T. M. Gassymov, in {\it Nekotorye Voprosy Eksp.
Teor. Fiz.}, (Elm, Baku, 1977), p. 3-27; Dokl. Akad. Nauk
Azerb. SSR {\bf 32} (6), 19 (1976).
\bibitem{29} T. M. Gassymov, in {\it Nekotorye Voprosy Teor.
Fiz.}, (Elm, Baku, 1990).   
\bibitem{30} T. M. Gassymov, Dokl. Akad. Nauk Azerb. SSR
{\bf 32}, 3 (1976); T. M. Gassymov and M. Y. Granowskii, Izv.
Akad. Nauk Azerb. SSR, Fiz. {\bf 1}, 55 (1976).
\bibitem{31} I. G. Kuleev, I. I. Lyapilin, A. A. Lanchakov,
and I. M. Tsidil'kovskii, Zh. Eksp. Teor. Fiz. {\bf 106}, 1205
(1994) [JETP {\bf 79}, 653 (1994)].
\bibitem{32} I. I. Lyapilin and K. M. Bikkin, in
{\it Proceedings of the 4th Russia Conference on Physics of
Semiconductors} (Novosibirsk, 1999), p. 52.
\bibitem{33} I. I. Lyapilin and K. M. Bikkin, Fiz. Tekh.
Poluprovodn. (St. Petersburg), {\bf 33}, 701 (1999)
[Semiconductors {\bf 33}, 648 (1999)].
\bibitem{34} I. G. Kuleev, A. T. Lonchakov, I. Yu. Arapova
and G. I. Kuleev, Zh. Eksp. Teor. Fiz. {\bf 114}, 191 (1998)
[JETP {\bf 87}, 106 (1998)].
\bibitem{35} S. S. Shalyt and S. A. Aliev, Fiz. Tverd. Tela
(Leningrad) {\bf 6} 1979 (1964).
\bibitem{36} S. A. Aliev, L. L. Korenblit, and S. S. Shalyt,
Fiz. Tverd. Tela (Leningrad) {\bf 7}, 1973 (1965).
\bibitem{37} I. N. Dubrovnaya and Yu. I. Ravich, Fiz. Tverd.
Tela (Leningrad) {\bf 8}, 1455 (1966).
\bibitem{38} V. I. Tamarchenko, Yu. I. Ravich,
L. Ya Morgovskii {\it et al.}, Fiz. Tverd. Tela (Leningrad)
{\bf 11}, 3506 (1969).
\bibitem{39} K. M. Bikkin, A. T. Lonchakov, and I. I.
Lyapilin, Fiz. Tverd. Tela (St. Petersburg) {\bf 42}, 202
(2000) [Phys. Solid State, {\bf 42}, 207 (2000)].
\bibitem{Bass} F. G. Bass, Yu. G. Gurevich, Soviet Journal of
Experimental and Theoretical Physics {\bf 52}, 175 (1967).
\bibitem{report} Yu G. Gurevich, O. L. Mashkevich, Physics
Reports, {\bf 181}, 327 (1989).
\bibitem{Mash1} Yu G. Gurevich, O. L. Mashkevich, Fiz.Tekh.
Poluprovodn. {\bf 15}, 659 (1981).
\bibitem{Mash2} Yu G. Gurevich, O. L. Mashkevich, Fiz.Tekh.
Poluprovodn. {\bf 15}, 1780 (1981).
\bibitem{Mash3} O. L. Mashkevich, Fiz.Tekh. Poluprovodn.
{\bf 15}, 1951 (1981).
\bibitem{Gur} L. E. Gurevich, T. M. Gassymov, Soviet Journal
of Solid State Physics (FTP) {\bf 9}, 106 (1967).
\end{thebibliography}
\end{document}